\documentclass[12pt,preprint]{aastex}
\usepackage{epsfig}
\usepackage{natbib}

\newcommand{\htwoo}{H$_2$O}

\newcommand{\cotwo}{\mbox{CO$_{2}$}}

\newcommand{\kms}{km~s$^{-1}$}

\newcommand{\mols}{molecules~s$^{-1}$}

\def\deg{\hbox{$^\circ$}}

\shorttitle{}
\shortauthors{Feldman et al.}

\begin{document}

\title{Ultraviolet Spectroscopy of Comet 9P/Tempel 1 with Alice/Rosetta
during the Deep Impact Encounter}

\author{Paul D. Feldman}

\affil{Department of Physics and Astronomy, The Johns Hopkins University\\ 
Charles and 34th Streets, Baltimore, MD 21218-2695}
\email{pdf@pha.jhu.edu}

\author{S. Alan Stern, Andrew J. Steffl, Joel Wm. Parker}

\affil{Space Science \& Engineering Division, \\
Southwest Research Institute, 1050 Walnut Street, Suite 400, Boulder, CO 80302}

\author{David C. Slater}

\affil{Space Science \& Engineering Division, \\ 
Southwest Research Institute, 6220 Culebra Road, San Antonio, TX 78238}

\author{Michael F. A'Hearn}

\affil{Department of Astronomy, University of Maryland, \\
College Park MD 20742-2421}

\author{Jean-Loup Bertaux}

\affil{Service d'A\'eronomie du CNRS, BP~3, F-91371 Verri\`eres-le-Buisson, France}

\author{Michel C. Festou\altaffilmark{1}}

\affil{Observatoire Midi-Pyr\'en\'ees, 14, avenue E.~Belin, F-31400 Toulouse, France}
\altaffiltext{1}{deceased May 11, 2005}

\pagestyle{myheadings}


\begin{abstract}

We report on spectroscopic observations of periodic comet 9P/Tempel~1
by the Alice ultraviolet spectrograph on the {\it Rosetta} spacecraft
in conjunction with NASA's Deep Impact mission. Our objectives were to
measure an increase in atomic and molecular emissions produced by the
excavation of volatile sub-surface material.  We unambiguously detected atomic oxygen emission from the quiescent coma but no
enhancement at the 10\% (1-$\sigma$) level following the impact.  We
derive a quiescent \htwoo\ production rate of $9 \times
10^{27}$~\mols\ with an estimated uncertainty of $\sim$30\%.  Our upper
limits to the volatiles produced by the impact are consistent with
other estimates.

\end{abstract}


\newpage
\section{INTRODUCTION}

The Deep Impact mission \citep{A'Hearn:2005a} successfully placed a 364
kg impactor onto the surface of comet 9P/Tempel~1 at a relative
velocity of 10.3~\kms\ on 2005 July 4 at 05:52:03 UT (as seen from
Earth).  The event was observed by cameras aboard the mother spacecraft
\citep{A'Hearn:2005b} and by a large number of Earth- and space-based
telescopes as part of an extensive campaign to study the comet prior
to, during, and in the course of several days following the impact
\citep{Meech:2005}.  Several instruments on board the {\it Rosetta}
spacecraft, which is on a trajectory for a rendezvous with comet
67P/Churyumov-Gerasimenko in 2014, were programmed to observe comet
Tempel~1 from June 28 through July 14, 2005.  During this period {\it
Rosetta} was located between 0.54 and 0.51~AU from the comet, which was
observed at a solar elongation angle varying from 93.4\deg\ to
88.1\deg\ and solar phase angle varying from 65.1\deg\ to 72.5\deg.  At
the time of impact these parameters were 0.531~AU, 91.1\deg, and
68.2\deg, respectively.  Perihelion, at 1.506~AU, occurred on July 5.
Imaging of the comet with the OSIRIS cameras on {\it Rosetta} has
been reported by \citet{Kuppers:2005} and \citet{Keller:2005}.

Alice was the only space-based spectrograph operating at wavelengths
below 2000~\AA\ at the time of Deep Impact.  The spectral region
spanned by Alice is rich is molecular and atomic emissions of the
primary volatile components of the cometary coma.  These include several
band systems of CO and the strongest resonance transitions of
the principal atomic species, C, H, O, N, and S, that have been
observed in many comets over the past 30 years using orbiting
observatories and sounding rockets \citep[e.g.,][]{Feldman:2004}.
Alice is also sensitive to emissions of oxygen and carbon ions.  All of
these species were expected to show an increase as a result of the
excavation of volatile sub-surface material by the impactor.

Because of the comet's relatively large distance from the Sun and its
low quiescent activity level, characteristic of Jupiter-family comets,
Alice was only able to unambiguously detect the quiescent coma in \ion{O}{1}~$\lambda$1304, from which a quiescent water production rate
of $\sim9 \times 10^{27}$~\mols\ is deduced.  \ion{H}{1}
Lyman-$\beta$ is also marginally detected above the interplanetary
background.  A slow secular decrease in the brightness of 
\ion{O}{1}~$\lambda$1304 with time was detected but we did not see any
significant increase (i.e., 10\%) as a result of the Deep Impact
event.  Our upper limits on the increase in these emissions (and those
of other species as well) are consistent with other reports of
\htwoo\ and CO production from the impact.

\section{INSTRUMENT}

Alice is a lightweight, low-power, imaging spectrograph optimized for
{\it in situ} cometary far-ultraviolet (FUV) spectroscopy. It is
designed to obtain spatially-resolved spectra in the
700-2050~\AA\ spectral band with a spectral resolution between 8 and
and 12~\AA\ for extended sources that fill its
0.05\deg\ $\times$~6.0\deg\ field-of-view.  Alice employs an off-axis
telescope feeding a 0.15-m normal incidence Rowland circle spectrograph
with a concave holographic reflection grating. The imaging
microchannel plate detector utilizes dual solar-blind opaque
photocathodes (KBr and CsI) and employs a two-dimensional delay-line
readout.  Details are given by \citet{Slater:2001}.  Following launch
on March 2, 2004, and instrument commissioning, Alice observed comet
C/2002~T7~(LINEAR) on April~30 and May~17, 2004.  These measurements
demonstrated the feasibility of remote ultraviolet observations from
{\it Rosetta}, and served as a baseline for the data obtained during
the Deep Impact campaign.

\section{OBSERVATION PROGRAM}

The Alice instrument observations during the {\it Rosetta} Deep Impact 
campaign consisted of a repeated series of five ``jailbar'' exposures 
stepped across Tempel~1.  Each step was shifted perpendicular to the 
slit length by 0.046\deg, which is slightly less than the slit 
width of 0.05\deg, providing some spatial overlap between steps.  
The five jailbar steps in each series therefore stepped the Alice 
boresight a total distance of 0.183\deg\ centered on Tempel~1.  
The exposure times were 26~minutes at the central step that pointed 
toward the target, and typically 19~minutes at the off-target 
positions.  The total exposure time on the last off-target point was 
70~minutes due to the longer dwell time to accommodate observations by 
the MIRO instrument.  This jailbar pattern was repeated almost 
continuously throughout the entire period from 2005 June 29 through 
July 14, except for three 3-hour interruptions for spacecraft reaction 
wheel offloads, and a few sets of dark exposures.  At the time of the
impact event, one set of jailbars was executed more quickly with
exposure times of 9~minutes at each pointing except the last off-target
point (coincident with the MIRO sub-millimeter boresight), which was 34
minutes.  We then returned back to the original set of nominal exposure
times for all subsequent jailbar observation sequences.

A calibration star, $\rho$~Leo (B1~Ib), was observed on 2005 
June 27 using the same jailbar pattern, though with different exposure 
times.  Longer single-pointing exposures of the star were made for 
additional calibration and to check spacecraft pointing stability.  The 
latter was checked by making exposures in time-tagged (``pixel-list'') 
mode to look for variations in the count rate from the star.  No 
variations were seen, implying the star was well-centered in the slit 
and that spacecraft pointing was accurate and stable relative to the 
width of the slit.


\section{SPECTRAL DATA}

The Alice slit consists of pixels that are 0.019\deg\ wide in the
spectral direction and 0.3\deg\ high perpendicular to the dispersion.
The width of the central region of the slit is 0.05\deg\ which
translates to a projected size of approximately $69,000 \times
416,000$~km$^2$ for a spatial pixel at the distance to the comet at the
time of impact.  Its large field-of-view, designed for {\it in situ}
coma spectroscopy, makes Alice sensitive to the extended hydrogen and
oxygen comae at the expense of emissions from the near-nucleus region.
For each of the jailbar positions data are accumulated in
two-dimensional arrays corresponding to wavelength and spatial axes.  A
data cube is produced by stacking these arrays with time as the third
dimension.  The data are displayed below as cuts along each of these
three axes.

Inspection of the data cube shows that most of the ambient cometary
emission occurs in the central row, row 15, of the spectral array.
This row, integrated over time and divided into two time periods: pre-
and post-impact, is shown in Fig.~\ref{spectra}.  Also shown are the
spectra from rows 13 and 12, which are displaced $8.30 \times 10^5$~km
and $1.24 \times 10^6$~km from the center of row 15, respectively,
projected on the sky.  The background is a combination of detector dark
counts and grating scattering of the interplanetary \ion{H}{1}
Lyman-$\alpha$ emission.  The latter is qualitatively similar to the
scattering function determined from pre-flight laboratory measurements
\citep{Slater:2001}.  The photocathodes were deposited on the
microchannel plate detector in such a way to leave a gap at
Lyman-$\alpha$ to avoid saturation of the detector.  This makes the
absolute calibration at this wavelength uncertain and we will limit our
discussion of hydrogen to Lyman-$\beta$.

\placefigure{spectra}

The figure shows the clear detection of cometary
\ion{O}{1}~$\lambda$1304, together with a possible enhancement of the
\ion{H}{1} emissions above the interplanetary background.  No other
feature is detected.  \ion{C}{1}~$\lambda$1657, usually the brightest
coma feature after Lyman-$\alpha$ and \ion{O}{1} \citep{Feldman:2004},
is not detected, but this may be in part due to the 3 times lower
instrumental effective area at 1657~\AA\ compared with 1304~\AA.  In
comet C/2002~T7~(LINEAR) on 2004 April 30 when the heliocentric
velocity dependent \ion{O}{1} and \ion{C}{1} fluorescence efficiencies
(evaluated at 1~AU) were similar to those for the Tempel~1
observations, the brightness of \ion{C}{1} as observed by Alice was
one-third that of \ion{O}{1}.  The absence of \ion{C}{1} emission in
Tempel~1 may reflect either a compositional difference or an effect of
the different observation geometries for the two comets.  There is no
significant difference between the pre-impact and post-impact spectra.
Upper limits to the expected \ion{C}{1} and CO emissions are given in
Table~\ref{tab1}.

\placetable{tab1}

The spatial distributions of \ion{O}{1}~$\lambda$1304 and \ion{H}{1}
Lyman-$\beta$ are shown in Fig.~\ref{spatial}, again separated into
temporal sums of the pre- and post-impact data.  The \ion{O}{1} is
clearly associated with the comet and serves as a verification of the
pointing of the spacecraft.  Lyman-$\beta$, however, is seen to be a
relatively minor enhancement to the interplanetary background.  This is
consistent with the coma modeling discussed below.  The variation of
the background Lyman-$\beta$ along the length of the slit is likely
due to an incomplete flat-field characterization of the detector
response.  Unfortunately, there were no deep sky observations to
characterize the interplanetary background, which had not been a
problem for the 2004 comet observations.  The general trend in the
variation of Lyman-$\beta$ along the slit can be seen from an average
of the two outer jailbar observations of $\rho$~Leo, as shown in
Fig.~\ref{Lbeta}.  The star was $\sim$23\deg\ from the comet.  While
the trend is similar to that seen in Fig.~\ref{spatial}, the star was
sufficiently bright that its point spread function gave a significant
contribution to rows 13--16 of the detector, even offset 0.09\deg\ from
the star, so it was not possible to use this profile to obtain a flat
field that would enable us to extract the cometary Lyman-$\beta$ with
any confidence.

\placefigure{spatial}
\placefigure{Lbeta}

The temporal variation of \ion{O}{1}~$\lambda$1304, shown as a daily
average, is given in Fig.~\ref{temporal}.  There is no immediate
increase following impact but this would not be expected as the
lifetime against photodissociation of \htwoo\ at 1.506~AU is about two
days and that for OH is about four days.  The high and low points seen
on days 2 and 3 following impact, respectively, are less than
2-$\sigma$ from the mean and are not statistically significant.  There
is also no indication, in the spectra obtained in the first few hours
after impact, of any emission at the wavelengths of the strongest CO
Fourth Positive bands, 1510~\AA\ (1,0), and 1478~\AA\ (2,0).  The
\ion{O}{1}~$\lambda$1304 emission does show a slow secular decrease
with time, and is consistent with the trend seen in the visible by
\citet{Schleicher:2006} and attributed to a slow decrease in the
comet's quiescent outgassing rate.

\placefigure{temporal}

\section{DISCUSSION}

\subsection{Quiescent Coma Models}

The H and O emissions can be modelled with a Haser model
\citep[see, e.g.][]{Combi:2004}, in which we account for saturation of
the solar flux with increasing column density along the line-of-sight
through the coma.  The model assumes radial outflow everywhere in the
coma, which is valid for the case of the long Alice slit, and uses
photodissociation rates appropriate to solar minimum from Table VI of
\citet{Budzien:1994}.  The mean outflow velocities used were 0.7~\kms\
for \htwoo, 19.6 and 6.0~\kms\ for H (from \htwoo\ and OH, respectively),
and 1.2~\kms\ for O.  Fluorescence efficiencies for \ion{O}{1} are
calculated following \citet{Dymond:1989} using solar fluxes appropriate
to solar minimum from \citet{Rottman:2001}, while for \ion{H}{1} solar
fluxes from SOHO/SUMER \citep{Lemaire:2002} are used.  The spatial profiles
for a model with an \htwoo\ production rate of $1 \times
10^{28}$~\mols, with the nucleus centered in row~15, are shown
overplotted on the data of Fig.~\ref{spatial}.  The model is probably
less reliable at distances beyond $10^6$~km from the nucleus because
the loss of H and O by charge exchange with solar wind ions is not
included in the model.  Comparing the model with the integrated
brightness in the 5 central rows leads to a derived quiescent
production rate of $9 \times 10^{27}$~\mols.  The uncertainty in this
result includes a 15\% absolute calibration uncertainty and, more
difficult to quantify, a 25\% uncertainty in the parameters used in the
Haser model.
 
This result may be compared with other estimates of the pre-impact
\htwoo\ production rate.  \citet{Kuppers:2005} using narrow-band OH
images recorded by the OSIRIS wide-angle camera on board {\it Rosetta},
found a quiescent value $Q$(\htwoo)~$= (3.4 \pm 0.5) \times
10^{27}$~\mols, while \citet{Schleicher:2006}, from ground-based OH
photometry, find $6 \times 10^{27}$~\mols.  Both of these use Haser
models similar to the one described above to interpret the observed
photodissociation product, OH.  Direct observations of water were made
in both the infrared, using long-slit spectroscopy, by
\citet{Mumma:2005}, and in the sub-millimeter, by the {\it Odin}
satellite, observing at 557 GHz \citep{Biver:2005}.  They report water
production rates of $10.4 \times 10^{27}$~\mols\ and $(8.5 \pm 1.5)
\times 10^{27}$~\mols, respectively.  Our result, within the
uncertainties described above, are in general accord with the infrared
and sub-mm observations of \htwoo.

\subsection{Gas Production Following Impact}

The upper limits on gas production derived from these data can then be
compared with expectations based on other observations.  For the amount
of \htwoo\ produced by the impact, \citeauthor{Kuppers:2005} give 
\mbox{$(1.5 \pm 0.5) \times 10^{32}$} molecules and
\citeauthor{Biver:2005} give $(1.4 \pm 0.35) \times 10^{32}$~molecules.  \citeauthor{Schleicher:2006} give an upper limit of $4 \times
10^{32}$~molecules.  Because their long slit did not include all of the
water produced by the impact, \citeauthor{Mumma:2005} only give an
effective production rate, which is twice their quiescent value. 

\citeauthor{Kuppers:2005} claim the total number of \htwoo\ molecules
produced to be about 15\% of the ambient number 
in the coma.  We don't observe \htwoo\ but two of its dissociation 
products.  In a time, $t$, the fraction of water dissociated into H and
OH (and eventually O) will be \mbox{(1 -- exp($-t/\tau$(\htwoo))}, where
$\tau$(\htwoo) at 1.506~AU is $1.88 \times 10^5$~s for solar minimum
conditions \citep{Budzien:1994}.  Thus, one day after impact we will
see $\sim$35\% of the 15\% increase in the amount of H, i.e., a 5\% increase
over ambient.  With the Alice slit the increase will be even less as
the fast moving H atoms ($\sim$20~\kms) will leave the area of the
projected slit in a matter of hours.  The increase in O will be even
smaller as two photodissociations are required and the lifetime of OH,
the source of 90\% of O in the coma \citep{Combi:2004}, is about twice
as long as that of \htwoo.  Promptly produced O~atoms, the
source of 10\% of the oxygen in the coma, will give a 1\% enhancement
after two days.  The S/N of the Alice data do not permit such a
detection.  CO and \cotwo\ are additional minor sources of oxygen
atoms but have photodissociation lifetimes greater than 10~days at
1.5~AU \citep{Huebner:1992}, and so will not produce a detectable change.

\subsection{Limits on Carbon Monoxide}

\citet{Feldman:2006} have modelled the images from the Solar Blind
Channel of the HST/ACS to derive a quiescent CO production rate of $4-6
\times 10^{26}$~\mols\ and a number of $1.5 \times 10^{31}$~CO
molecules produced by the impact.  Modeling the upper value of CO
quiescent outgassing in terms of radial outflow from the nucleus, we
can estimate the expected CO Fourth positive emission in the central
spatial pixel to be 0.024 rayleighs for the strongest (1,0) and (2,0)
bands.  The change in CO resulting from the impact will give
an increase in brightness of 0.004 rayleighs in these bands, the
low value being a consequence of the large FOV of the Alice slit and
the concentration of the produced gas within a few thousand~km of the
nucleus.  These values are well below the 3-$\sigma$ upper limits given
in Table~\ref{tab1}.  Because of the long photodissociation lifetime of
CO, the atomic oxygen and carbon produced from CO will again give a
negligible contribution to either the \ion{O}{1}~$\lambda$1304 or
\ion{C}{1}~$\lambda$1657 brightness.

\section{CONCLUSION}

Observations of periodic comet 9P/Tempel~1 were made with the Alice
ultraviolet spectrograph on the {\it Rosetta} spacecraft from six days
before through ten days following the Deep Impact encounter with the
comet.  \ion{O}{1} emission from the quiescent comet was clearly
detected and an \htwoo\ production rate of $9 \times 10^{27}$~\mols,
with an estimated uncertainty of $\sim$30\%, was derived from the
measured \ion{O}{1} brightness profile.  Any enhancement either
immediately following the impact or in the subsequent ten days was less
than 10\% (1-$\sigma$).  Our upper limits to the volatiles produced by
the impact are consistent with estimates from other observations.

\begin{center}{\bf ACKNOWLEDGMENTS}\end{center}

We thank the ESA Rosetta Science Operations Centre (RSOC) and Mission
Operations Center (RMOC) teams for their expert and dedicated help in
planning and executing the Alice observations of comet Tempel~1.  We
also thank Claudia Alexander (JPL), Kristin Wirth (ESTEC), Detlef
Koschny (ESTEC), Paolo Ferri (ESOC), Maarten Versteeg (SwRI), and
Gerhard Schwehm (ESTEC) for their continued support of Alice.  The
work at Johns Hopkins University was supported by NASA's Jet Propulsion
Laboratory through a sub-contract from Southwest Research Institute.


\clearpage

\renewcommand\baselinestretch{1.1}%

\begin{deluxetable}{r*{5}{r@{}c@{}l}}
  \tablewidth{0pt}
  \tabletypesize{\footnotesize}
  \tablecaption{24-hour Averaged Upper Limits
    \label{tab1}}
  \tablehead{\colhead{} & \multicolumn{15}{c}{Days From Impact} \\
    \colhead{Feature} & \multicolumn{3}{c}{--1 to 0} &
    \multicolumn{3}{c}{0 to 1} & \multicolumn{3}{c}{1 to 2} &
    \multicolumn{3}{c}{2 to 3} & \multicolumn{3}{c}{3 to 4}} 
  \startdata

  CO 1510 \AA\ & & $<$ & 0.83 & & $<$ & 0.53 & & $<$ & 0.66 & & $<$ & 0.54
  & & $<$ & 0.79 \\

  \ion{C}{1}~1561 \AA\ & & $<$ & 0.51 & & $<$ & 0.42 & & $<$ & 1.13 & & $<$ & 0.55
  & & $<$ & 1.27 \\

  \ion{C}{1}~1657 \AA\ & & $<$ & 0.86 & & $<$ & 1.21 & & $<$ & 1.19  & & $<$ & 1.00
  & & $<$ & 1.00 \\
  
  \enddata
  
  \tablecomments{Upper limits are 3-$\sigma$ and are given in rayleighs, extracted from detector row 15 using observations made in the central
jailbar position.}

\end{deluxetable}

\clearpage
\renewcommand\baselinestretch{1.2}%

\clearpage 
\begin{center}{\bf FIGURE CAPTIONS}\end{center}

\figcaption{Pre- and post-impact spectra from jailbar C.  In both panels
rows 15, 13, and 12 are shown, top to bottom, dispaced by 
0.05~rayleighs~\AA$^{-1}$ for clarity.  Rows 13 and 12 are displaced 
$8.30 \times 10^5$~km and $1.24 \times 10^6$~km from the center of
row 15, respectively, projected on the sky.  The total exposure times 
are 56,608~s (pre-impact) and 111,402~s (post-impact).  \label{spectra}}

\figcaption{Spatial profiles of \ion{O}{1}~$\lambda$1304 and 
\ion{H}{1}~Lyman-$\beta$ derived from the same data shown in
Fig.~\ref{spectra}.  The observed Lyman-$\beta$ is primarily from
interplanetary hydrogen rather than the comet.  Statistical error
bars (1-$\sigma$) are shown.  The result of a Haser model for an
\htwoo\ production rate of $1 \times 10^{28}$~\mols, with the nucleus
centered in row~15, is indicated by $\times$ in the pre-impact
profiles. \label{spatial}}

\figcaption{Spatial profile of interplanetary \ion{H}{1}~Lyman-$\beta$
derived from the outer jailbars of the $\rho$~Leo observation.  The
total integration time is 5,800~s.  Statistical error bars (1-$\sigma$)
are shown.  Rows 13 to 16 are contaminated by light from the star.
\label{Lbeta}}

\figcaption{Daily average brightness of the observed \ion{O}{1}~$\lambda$1304
emission.  The data are the mean of rows 14 to 16 and the error bars are
1-$\sigma$ in the observed counts.  \label{temporal}}

\setcounter{figure}{0}
\clearpage

\begin{figure*}
\begin{center}
\epsscale{0.85}
\rotatebox{0.}{
\plotone{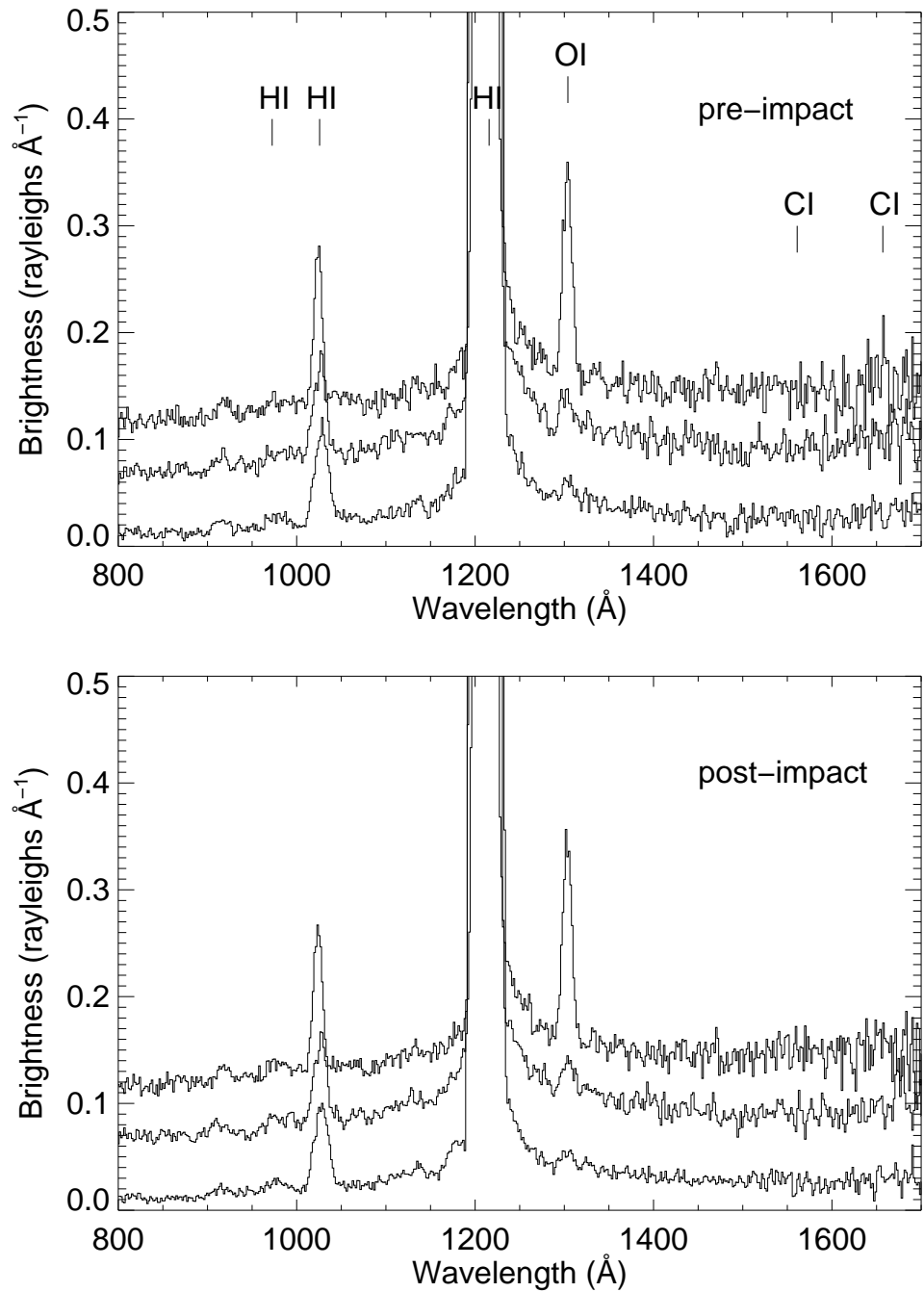}}
\caption{Pre- and post-impact spectra from jailbar C.  In both panels
rows 15, 13, and 12 are shown, top to bottom, dispaced by 
0.05~rayleighs~\AA$^{-1}$ for clarity.  Rows 13 and 12 are displaced 
$8.30 \times 10^5$~km and $1.24 \times 10^6$~km from the center of
row 15, respectively, projected on the sky.  The total exposure times 
are 56,608~s (pre-impact) and 111,402~s (post-impact).}
\end{center}
\end{figure*}

\begin{figure*}
\begin{center}
\epsscale{1.0}
\rotatebox{0.}{
\plotone{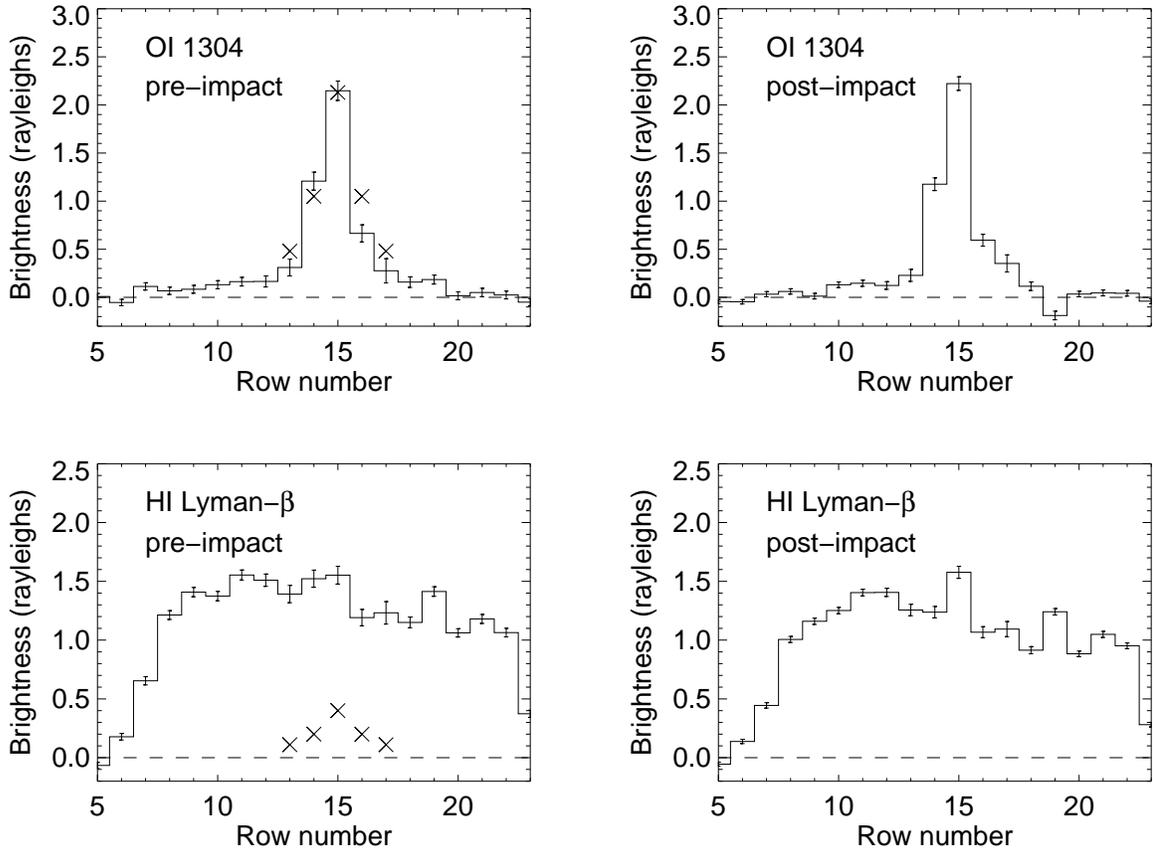}}
\caption{Spatial profiles of \ion{O}{1}~$\lambda$1304 and 
\ion{H}{1}~Lyman-$\beta$ derived from the same data shown in
Fig.~\ref{spectra}.  The observed Lyman-$\beta$ is primarily from
interplanetary hydrogen rather than the comet.  Statistical error
bars (1-$\sigma$) are shown.  The result of a Haser model for an \htwoo\ production
rate of $1 \times 10^{28}$~\mols, with the nucleus centered in row~15,
is indicated by $\times$ in the pre-impact profiles. }
\end{center}
\end{figure*}

\begin{figure*}
\begin{center}
\epsscale{0.6}
\rotatebox{0.}{
\plotone{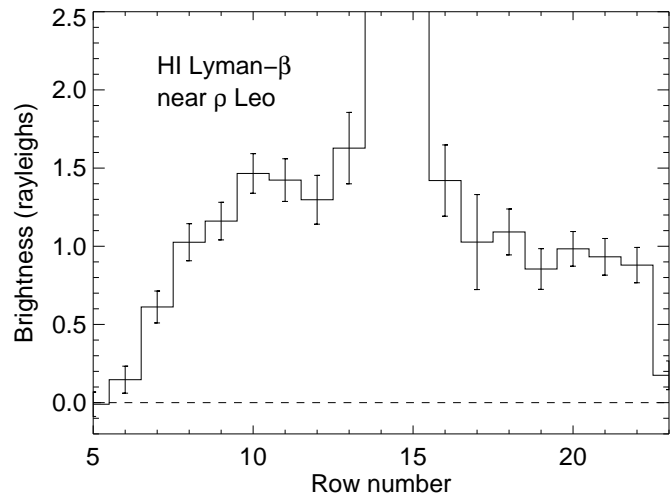}}
\caption{Spatial profile of interplanetary \ion{H}{1}~Lyman-$\beta$
derived from the outer jailbars of the $\rho$~Leo observation.  The
total integration time is 5,800~s.  Statistical error bars (1-$\sigma$)
are shown.  Rows 13 to 16 are contaminated by light from the star. }
\end{center}
\end{figure*}

\begin{figure*}
\begin{center}
\epsscale{1.0}
\rotatebox{0.}{
\plotone{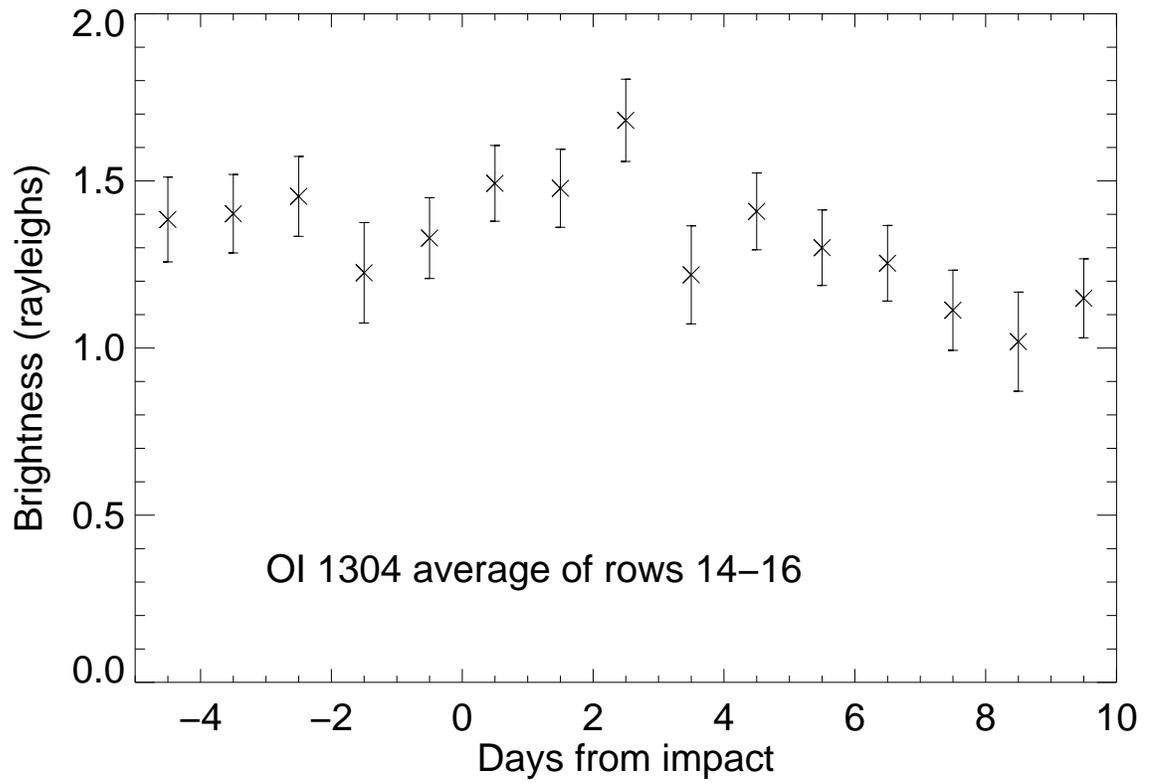}}
\caption{Daily average brightness of the observed \ion{O}{1}~$\lambda$1304
emission.  The data are the mean of rows 14 to 16 and the error bars are
1-$\sigma$ in the observed counts.}
\end{center}
\end{figure*}

\end{document}